# Lossless Microarray Image Compression by Hardware Array Compactor

Anahita Banaei, Shadrokh Samavi and Ebrahim Nasr Esfahani

**Abstract:** Microarray technology is a new and powerful tool for the concurrent monitoring of a large number of gene expressions. Each microarray experiment produces hundreds of images. Each digital image requires a large storage space. Hence, real-time processing of these images and transmission of them necessitates efficient and custom-made lossless compression schemes. In this paper, we offer a new architecture for the lossless compression of microarray images. In this architecture, we have used dedicated hardware for the separation of foreground pixels from background ones. By separating these pixels and using pipeline architecture, a higher lossless compression ratio has been achieved as compared to other existing methods.

**Keywords:** DNA, Microarray, Image processing, Image compression, Pipeline.

## 1 Introduction

Genetics has been essential to all biomedical and biological research fields in recent years. Considering various applications of this science and its vital role in the discovery and diagnosis of various diseases, at the time being, extensive research is being done on new technologies and methods in this area.

Chromosomes are the long threads inside the nucleus of cells of all organisms. Each chromosome is composed of two types of large organic molecules (macromolecules) called proteins and nucleic acids. The nucleic acids are deoxyribonucleic acid (DNA) and ribonucleic acid (RNA). The genetic information of organisms is encoded in the structure of DNA. DNA consists of two long chains of subunits twisted around one another to form a double-stranded helix. The subunits of each strand are nucleotides, each containing any one of four chemical constituents called bases. The four bases in DNA are Adenine (A), Thymine (T), Guanine (G), and Cytosine (C). The bases in the double helix are paired. At any position on the paired strands of a DNA molecule, if one strand has an A, the partner strand has a T, and if one strand has a G, the partner strand has a C. the pairing between A-T and C-G is said to be complementary [1,2]. Figure 1 shows the double helix structure of a typical DNA.

The genetic information of an organism is transmitted from cell to cell during development by the accurate replication of the sequence of bases in nucleic acids due to the precise base pairing in double-stranded nucleic acids. When the two strands of a parental double helix of DNA separate, the base sequence of each parental strand can serve as a template for the synthesis of a new complementary strand.

Genes are encoded in the sequence of chemicals that make up DNA. A particular gene is said to express into a protein when it codifies that. One of the genetic research methods is to study the gene expression processes. Scientists can achieve a lot of information about the manner of activity of each gene, the structure of a protein made in a specific cell or detecting a particular gene in an organism by studying various phases of gene expression. Because of the large number of genes in an organism (for example a human body has 30000 to 40000 genes), the number of experiments that are required to fully characterize an organism by means of gene expressions is huge. Traditional methods of laboratory experiments required years of investigation to characterize a disease [3,4].

Microarray analysis is a recently developed technique, which allows study and classification of genes in a much shorter time than ever before. Nowadays microarray technology has turned to be one of the main tools for genetic researches. A microarray experiment can monitor the expressions of thousand of genes simultaneously [5].

The substrate of a microarray consists of a piece of glass, or sometimes a silicon chip, similar to a microscope slide. Thousands of patches of single-stranded DNAs which are called probes are fixed (spotted) onto this substrate by a robotic arrayer. A typical microarray is a 2×4cm membrane or a microscope slide with a probe diameter of 75-100µm and a 150µm distance between probes. The location and sequence of each patch of DNA are known. A leading use of DNA is in determining which subset of a cell's genes are expressed, or are actively making proteins under certain conditions such as exposure to a drug, toxic material, or malignancy [6].

Microarray technology is based on the ability of complementary base pairing of the nucleic acids. Probes are single strands with known sequences and are used as a template to identify the unknown agents. Target DNA mixture is then washed over onto the chip to allow base pairing that means only highly complementary sequences will remain bound to their pairs. Single strands in the target mixture may come for instance from healthy and cancerous cells. The goal could be to identify genes that are responsible for malignancy [7].

The target DNA mixture is labeled with different fluorescents dyes to distinguish between DNA originating from different experimental conditions. For instance, DNA from blood cancer cells may be labeled with the red fluorescing dye and that from normal blood cells with the green fluorescing dye. Then a laser scanner at two wavelengths or channels scans the microarray, one channel for each dye. Fluorescent intensity corresponding to each dye is recorded separately for each spot on the array. The products resulting from the array-scanning process are two fluorescent intensity images. These images are superimposed for each spot to arrive at the actual gene expression pattern for the cancerous blood cells. Array of spots fluorescing purely red present genes expressed only under cancer condition, while those that are pure green correspond to gene expressed only under normal conditions. Genes that are expressed under both conditions will appear as spots of varying degrees of yellow. The intensity ratio for each probe or spot is proportional to the relative abundance of DNAs of the two different samples [1]. A sample gray scale microarray image is shown in Fig. 2 [8]. The size of the image is 256×256 pixels and each spot is about 12×12 pixels. As shown in this figure, microarray images have regular structures and spots normally have circular shapes. In microarray images, spots form the foreground of the image.

## 2  Microarray Image Compression

Each microarray experiment produces thousands of images. Microarray images are large in size and also each experiment is costly. Various organizations share their microarray databases. For efficient storage and sharing large number of these images, image compression is essential [10,13].

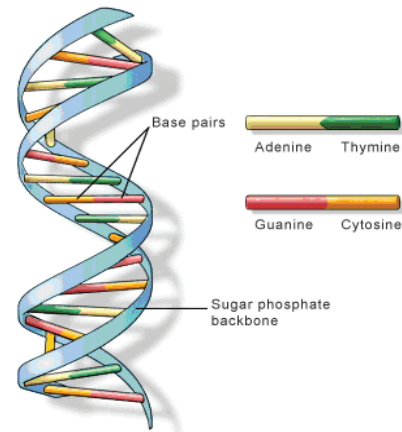

**Fig. 1** Structure of a DNA

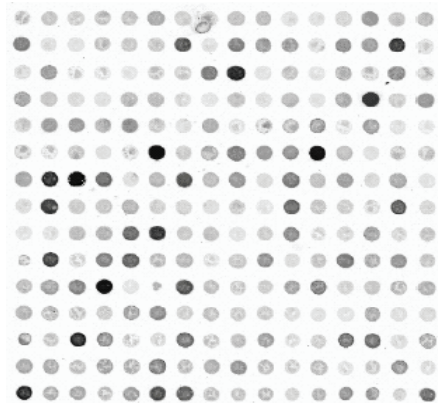

**Fig. 2** A sample gray scale microarray image [8]

A compression algorithm takes an input $\chi$ and generates a representation $\chi_c$ that requires fewer bits. There should also b a reconstruction algorithm that operates on the compressed representation $\chi_c$ to generate the reconstruction $\chi_r$. Based on the requirement of reconstruction, data compression schemes can be divided into two broad classes: lossless compression schemes, in which $\chi$ are identical to $\chi_r$, and lossy compression schemes, which generally provide much higher compression than lossless compression but allow $\chi_r$ to be different from $\chi$ [9]. "Huffman Coding" [10] and RLE (Run Length Encoding) [11] are two well known lossless compression schemes and "Vector Quantization" and "Transform Coding" are among lossy schemes [9]. Discrete Cosine Transform (DCT) and Discrete Wavelet Transform (DWT) are two of the techniques used for transform coding. These techniques are not lossy by nature but a quantization that is used afterward makes them lossy. Quantization simply reduces the number of bits needed to store the transformed coefficients by reducing the precision of those values. Since this is a many-to-one mapping, it is a lossy process and is the main source of compression in an encoder.

Compression of microarray images could be done using lossless or lossy methods. It is obvious that compression ratio of lossless methods is smaller than of lossy methods. For biomedical applications in which the pixel values have important information, lossless methods are the more appropriate choice. In microarray images, the intensity level of each spot contains important information about the gene expression. Loosing any part of this information could ruin the experiment and could result in completely different outcomes. Thus, for preserving the intensity of spots and their size by the compression algorithm, lossless schemes are more suitable. Until now, several methods have been proposed for the microarray image compression. Most of them are software implemented while a number of these methods have been realized by hardware.

Jornsten *et al.* proposed a software compression method called SLOCO (Segmented LOCO) which is based on LOCO (LOw COmplexity) that is used in JPEG2000 standard [12,13]. Hua *et al.* proposed a method called BASICA (Background Adjustment, Segmentation, Image Compression and Analysis). In BASICA object-based EBCOT (Embedded Block Coding Optimized Truncation) is used for lossless image compression [14]. Lonardi and Luo proposed a software called "Microzip" that after necessary processes, compress foreground values in a lossless form and background values by lossy means [15]. Faramarzpour *et al.* proposed a software lossless compression method. This method is based on the inherent property of microarray images which is the circular shape of spots. The idea in this method is to convert the 2D structure of the image into a 1D sequence which can scan the image in a highly correlated manner while preserving its spatial continuity [16]. Their method considers worst case scenarios and hence, does not come up with high compression ratios. A lossy method is proposed in [17] using a software routine. In that method the spots in the microarray image are extracted and a circle region is superimposed onto each of the spots. Then a *circle to square* (C2S) transform is performed to transform the area inside the circle of each spot to a corresponding square shaped image. The image is constructed by tiling the C2S transformed images together. Then the image is divided into 8×8 blocks and DCT is applied to each block. The transformed blocks are quantized and the image is coded by variable length coding.

A hardware method has been applied in [18] for lossless compression of microarray images. The input image which is inserted row-by-row to the pipeline architecture, is processed in the first stages and is compressed and transmitted in the final stages of the pipeline. Karimi *et al* proposed in [19] two pipeline architectures for the microarray image compression. In the first architecture, "Pseudo-RLE" method is applied and in the second architecture, "Residual Huffman Coding" method is used for the compression of images. In the residual Huffman coding method, codes are calculated for the difference between every two neighboring pixels. Samavi et al [20] proposed another hardware method based on pipeline architecture for microarray image compression. The image is first processed and through morphological operations noise and very small spots are eliminated. For compression purposes they process the pixels of the image in a raster scan order. Then a predictive coding algorithm is used to produce a residual sequence for each raster scan. Finally, Huffman coding is performed on the small-magnitude residual sequences.

In this paper we propose a hardware method which can compress microarray images with close to real time speeds. The achieved compression ratio is even better than some of the software routines that are in the literature. The proposed hardware could be used for other applications such as implementation of run length encoding.

In section 3 of the paper the proposed architecture is explained. Simulation results are presented in section 4 and concluding remarks are offered in section 5.

## 3 The Proposed Architecture

In this section, the proposed architecture is presented for the compression of microarray images. The block diagram of the structure is shown in Fig. 3. In this method, image data enters the circuit in a row-by-row manner. After the arrival of one row of the image, the foreground and background pixels are separated from each other. Foreground of a microarray image is the collection of pixels whose values are non-zero (spot pixels) and hence, background is the pool of pixels with zero values. Because of zero values of the background pixels, it is enough to store the coordinates. For this reason, corresponding bitmap of each row is calculated: for each pixel of the row which has zero value a "0" bit and for those with non-zero values, a "1" bit is inserted in the corresponding bitmap. Consequently, in the resulting bitmap, bits corresponding to background are 0 and those corresponding tosspots or foreground are 1. Here a new method is used for compressing the bitmap of each row. In this method, indices of the starting of strings 0's and 1's in the bitmap are extracted. Data that should be transmitted for each row is the extracted indices instead of the whole bitmap. At the receiver end the original bitmap can be reconstructed using these indices. A dedicated hardware is used to acquire the indices, which will later be explained in details.

On the other hand, the foreground or the spot values, which have corresponding 1's in the bitmap, are separated from the background. It is also done by the dedicated hardware that is used to acquire the indices of the starting of strings in the bitmap. Details of this hardware will later be explained. Extracted foreground values can be compressed using various compression methods. Here, residual Huffman coding is used. For this purpose residual values of neighboring foreground pixels are calculated and coded using Huffman coding method.

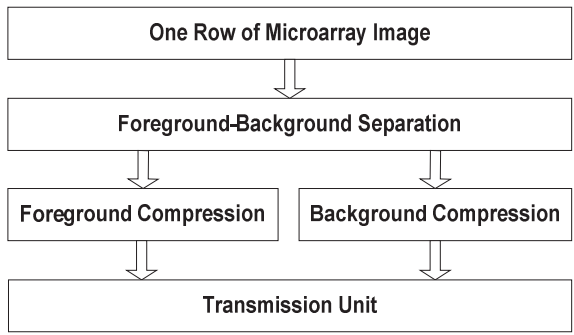

**Fig. 3** Block diagram of the proposed architecture

Pipeline stages for foreground compression are shown in Fig. 4. It is assumed that each row of the input image contains 256 pixels and the value of each pixel is presented by an 8-bit value. One row of image is inserted to the pipeline in each cycle of pipeline clock. The bitmap of the input row is calculated in the first stage of the pipeline shown in Fig. 4. The output of this stage is a 256-bit bitmap in which one bit is assigned to each pixel of the inserted row. If the value of a pixel is zero, then the corresponding bitmap position will be 0, otherwise a 1 appears in the bitmap. For calculating the bitmap corresponding to the input row, 8 bits of each pixel value are "OR"ed with each other. For this reason 256 8-input OR gates are used and thus the operation of producing bitmap is done in parallel.

The calculated bitmap contains strings of 1's and 0's. In the next stage of the pipeline that is shown in Fig. 4, the starting locations of these strings are obtained. By "XOR"ing the neighboring values, in the beginning of a new string where there is a 0 to 1 or 1 to 0 transition the output will be 1. In this way for the inner pixels of the strings where the neighboring values are the same, the output will be 0. Because each row begins with background or zero values, the first string of the bitmap is always a 0 string and following strings are alternatively 1's and 0's. Thus knowing only the place of the beginning of each string is enough and the type of the strings are easily distinguishable. Hence, the indices of elements that are 1 in the output vector of the second stage are the beginning position of the strings of the bitmap. We present a new hardware called Compression Unit (C.U.) for obtaining these indices. Details of a C.U. are shown in Fig. 5. Inputs of a C.U. are X and Y vectors. This hardware compresses the elements of X by eliminating those elements of X which have corresponding zero in the Y vector.

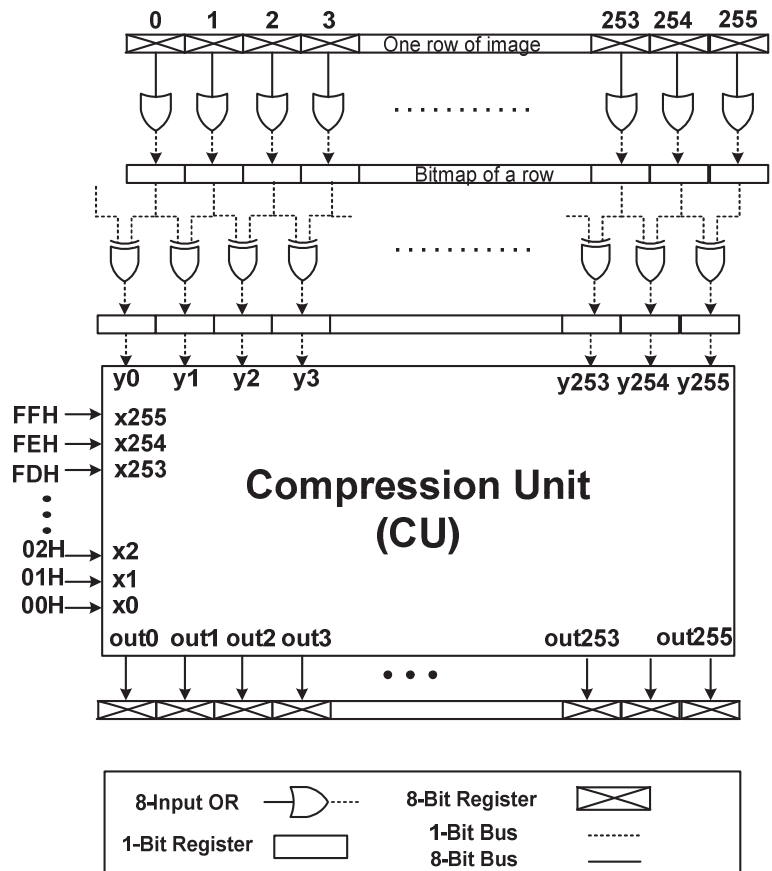

**Fig. 4** Pipeline stages for the background compression

The designed C.U. unit is composed of units which we call them the Routing Units (R.U.). Details of an R.U. are shown in Fig. 6. Also shown in that figure are all possible combinations of control signals, Cin1, Cin2, and their corresponding routing action. For larger or smaller input vectors, the C.U. could be scaled up or down. Since the rows of the input image have 256 pixels, the C.U. that is shown in Fig. 5 has 256-element input vectors.

Compression of the bitmap corresponding to the input row is done by using a compression unit (C.U.). This is performed by assigning 0 through 255 to the X vector. Vector Y is the output of the second pipeline stage in Fig. 4. The output of C.U. is a 256-element vector whose elements are the indices of beginning of the bitmap strings. Only a small numbers of the elements, towards the beginning of the output vector, have significant values. This is because the number of strings in a row of a bitmap is much smaller than 256.

strings. These sparse positions are fed into the CU and the output is a dense and ordered group of indices corresponding to the starting positions of the strings of a row. Figure 7 shows an example for the routing operation of the CU. The input to the CU is, for example, 00100101. The CU is to place the indices of the 1's at its input in a dense manner at its output.

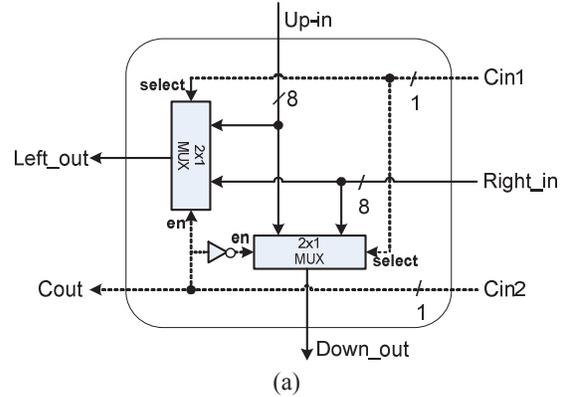

(a)

| Control line Cin1 | Control line Cin2 | Transference |
|---|---|---|
| 0 | 0 | Up_in →Down_out |
| 1 | 0 | Right_in →Down_out |
| 0 | 1 | Up_in →Left_out |
| 1 | 1 | Right_in →Down_out |

**Fig. 6** Details of the R.U. unit (a) structure (b) various states of control signals

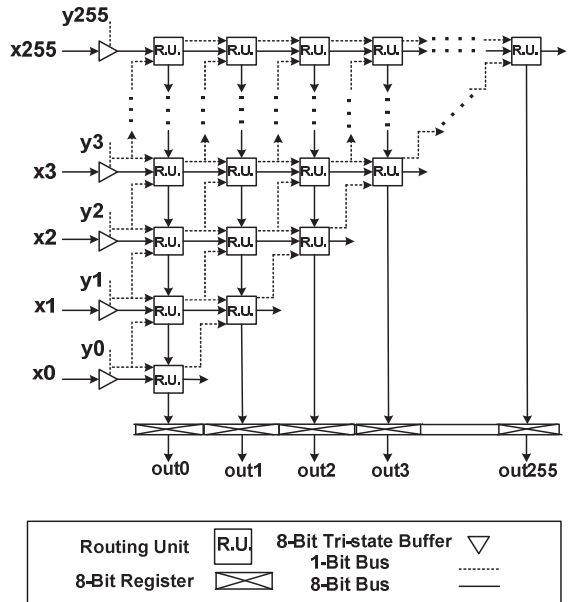

**Fig. 5** Internal structure of the compression unit (CU).

**Table 1** An example for different processes performed on a row of bitmap.

| A partial bitmap of a row | 0 0 1 1 1 1 1 0 0 1 1 1 1 1 0 0 |
|---|---|
| Starting position of the strings | 0 0 1 0 0 0 0 1 0 1 0 0 0 0 1 0 |
| Corresponding index of each position | 0 1 2 3 4 5 6 7 8 9 A B C D E F |
| Out put of the CU | 2 7 9 E 0 0 0 0 0 0 0 0 0 0 0 0 |

Every row starts with a string of zeros; therefore, it is not necessary to save the index value for the beginning of these strings. Hence, the first string that we encounter in a row is a string of 1's. Furthermore, every row ends with another string of zeros but the index value of start of that string needs to be saved. By placing a zero at the input of the first XOR at the left in Figure 4 we consider continuity between consecutive lines which creates higher compression ratio. In effect, the image is scanned in a raster scan manner.

Table 1. shows different steps of the pipeline operations on a row of the bitmap. In one step the starting positions of different strings of 1's and 0's are found. We are only interested in the index value of these starting positions. We know that the first index in a row belongs to a string of 1's and from there on they alternate between 0 and 1

Figure 8(a) shows an example with four spots in an 18×18 microarray image. All of the non-zero pixels belong to the spots and are considered as foreground. Each pixel has an eight-bit grayscale value. Therefore, to store this image 18×18×8 bits are required. Figure 8(b) shows the bitmap of the image where the 1's are for the foreground pixels and 0's correspond to the background pixels. Figure 8(c) shows the starting location of the strings of 0's and 1's in the bitmap. Only when a new string starts we see a 1 and everywhere else we have 0's. Figure 8(d) demonstrates how the indices of these starting positions are packed together. Only these values are required to be stored or transmitted in order to be able to reconstruct the bitmap. Another part of the hardware is responsible to get the foreground pixels and using a separate CU compresses them. Figure 8(e) shows the resulted compressed foreground. These values are further compacted using Hoffman coding. Using a reconstructed bitmap the compressed foreground pixels can be formed again.

The required number of bits to store the foreground and bitmap information without any further compression scheme is $(8\times24\times4)+(8\times54)$. This is because there are 24 pixels in each of the 4 spots. To indicate an all-zero row an eight-bit code can be designated. In this way 48 starting indices and 6 all-zero rows are to be stored. This shows that even without any type of variable length coding a compression ratio of 2.16 is achieved. Compression of the foreground pixels would further increase the compression ratio.

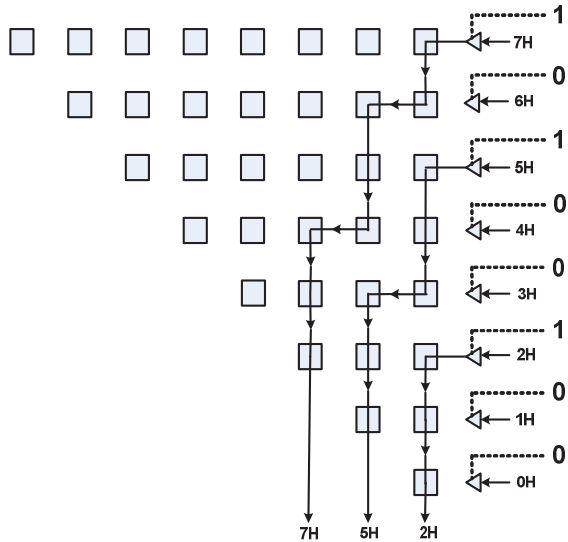

**Fig. 7** An example of the routing operation in a Compression Unit.

Separation and compression of foreground is done in parallel with compression of the background. In order to separate the foreground from the background, yet another C.U. is used. After a row of the image enters the pipeline and its bitmap is produced, the original row and its corresponding bitmap are inserted to the C.U. block. The hardware used for the foreground compression is shown in Fig. 9. This part of the circuit works in parallel with the second stage of the pipeline shown in Figure 4. The output of the C.U. block of Figure 9 is a vector whose elements are the foreground pixels of the row, bunched up towards the left, while keeping their order. Then a predictive technique is used that capitalizes on inter-pixel spatial redundancy. To do so, we predict the next pixel based on the values of the previous neighboring pixels. This is done by computing the residual pixel which is the difference between the two neighboring pixels. Finally, we losslessly compress the residual data, using Huffman coding. Therefore, the information of a row has been de-correlated which causes the residual to have lower entropy.

## 4 Simulation Results

In this section we discuss the simulation results of the proposed compression architecture. We used a range of standard microarray images. Three of these images have been used by MicroZip and 14 images are from ISREC set. These images can be accessed from references [21] and [22] respectively. Using the mentioned predictive method, an average compression ratio of 3.89:1 is obtained for compression of the foreground values. Even though, the method is lossless, we obtained high compression ratio compared to the conventional lossy and lossless methods.

Table 2 shows a quantitative comparison of the results of our proposed algorithm with some other algorithms. The hardware proposed in this paper was able to increase the compression ratio up to 2.17 times those presented in the literature.

**Table 2** Compression ratio of our method compared to some others

| Average compression ratio | Compression method |
|---|---|
| 2.46 | MicroZip [15] |
| 1.83 | SLOCO [13] |
| 2.04 | BASICA [14] |
| 2.13 | Reference [16] |
| 1.79 | Reference [20] |
| 3.89 | Our method |

Most of the routines mentioned in Table 1 are software based. Reference [20] is a hardware implementation and speed of processing is of main interest not the compression ratio. In this paper we achieved higher compression ratio than the other references due to the following factors. First the foreground was separated from the background hence better spatial correlations have been achieved. Secondly, the new compression unit that is introduced here has enabled the system to compact the data more densely.

## 5 Conclusion

In this paper, we presented a new lossless compression method for microarray images. By statistical studies on microarray images, we concluded that for increasing compression ratio we should separate the foreground pixels form those of the background. For this reason a new hardware was designed and used that could extract out the foreground values and place them next to each other in an orderly fashion. The proposed method has increased the compression ratio significantly. Comparison of the results of this method with those of other lossless and lossy methods proves the efficiency of this method.

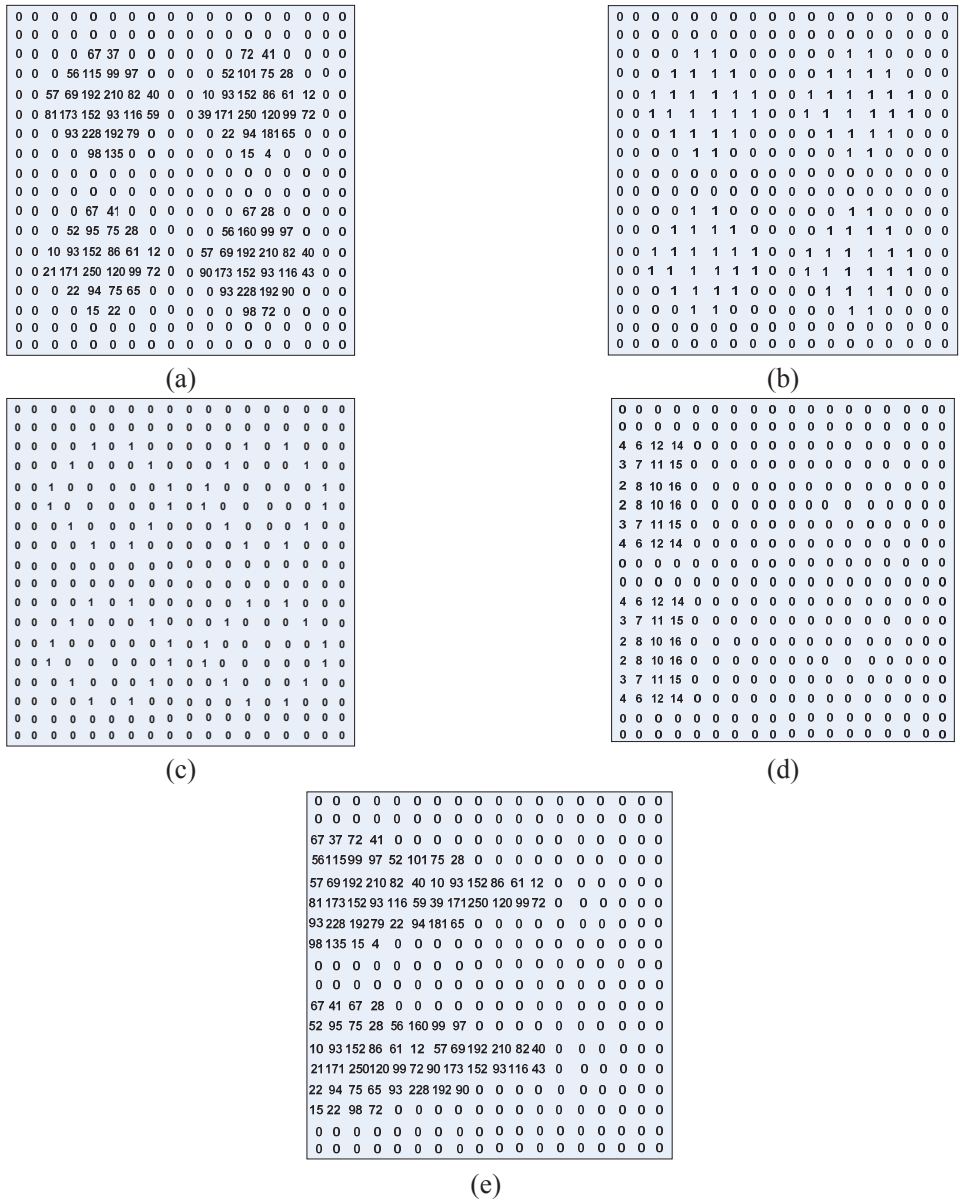

**Fig. 8** Different steps of the suggested compression method shown for four spots.

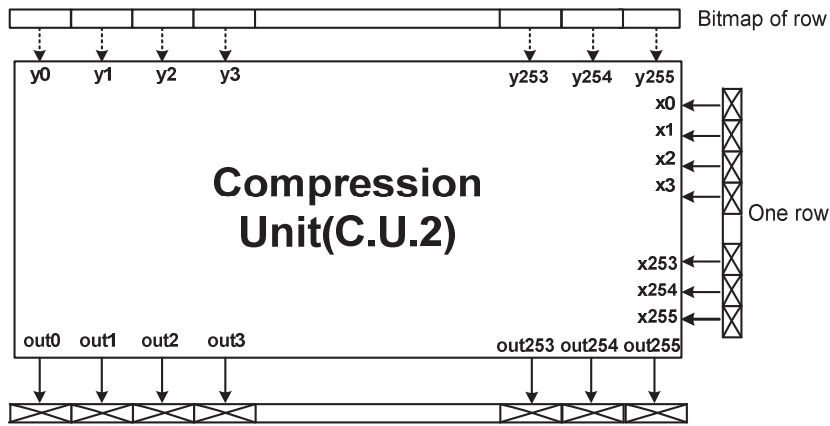

**Fig. 9** Circuitry for the foreground compression